\documentclass[conference]{IEEEtran}
\IEEEoverridecommandlockouts
\usepackage{cite}
\usepackage{amsmath,amssymb,amsfonts}
\usepackage{algorithmic}
\usepackage{graphicx}
\usepackage{textcomp}
\usepackage{xcolor}
\usepackage{cite}
\usepackage[subrefformat=parens]{subcaption}
\usepackage{multirow}
\usepackage{url} 
    \def\BibTeX{{\rm B\kern-.05em{\sc i\kern-.025em b}\kern-.08em
    T\kern-.1667em\lower.7ex\hbox{E}\kern-.125emX}}
\begin{document}

\makeatletter
\newcommand{\linebreakand}{%
  \end{@IEEEauthorhalign}
  \hfill\mbox{}\par
  \mbox{}\hfill\begin{@IEEEauthorhalign}
}
\makeatletter

\title{PrometheusFree: Concurrent Detection of Laser Fault Injection Attacks in Optical Neural Networks 
}

\author{\IEEEauthorblockN{Kota Nishida}
\IEEEauthorblockA{The University of Osaka \\
Suita, Japan \\
nishida.kota@ist.osaka-u.ac.jp}
\and
\IEEEauthorblockN{Yoshihiro Midoh}
\IEEEauthorblockA{The University of Osaka \\
Suita, Japan \\
midoh@ist.osaka-u.ac.jp}
\and
\IEEEauthorblockN{Noriyuki Miura}
\IEEEauthorblockA{The University of Osaka \\
Suita, Japan \\
nmiura@ist.osaka-u.ac.jp}
\linebreakand
\IEEEauthorblockN{Satoshi Kawakami}
\IEEEauthorblockA{Kyushu University \\
Fukuoka, Japan \\
kawakami@ed.kyushu-u.ac.jp}
\and
\IEEEauthorblockN{Alex Orailoglu}
\IEEEauthorblockA{University of California, San Diego \\
La Jolla, USA \\
alex@cs.ucsd.edu}
\and
\IEEEauthorblockN{Jun Shiomi}
\IEEEauthorblockA{The University of Osaka \\
Suita, Japan \\
shiomi-jun.ist@osaka-u.ac.jp}
}

\maketitle
\begin{abstract}
Silicon Photonics-based AI Accelerators (SPAAs) have been considered as promising AI accelerators achieving high energy efficiency and low latency. While many researchers focus on improving SPAAs’ energy efficiency and latency, their physical security has only recently received attention. While it is essential to deliver strong optical neural network inferencing approaches, their success and adoption are predicated on their ability to deliver a secure execution environment. Towards this end, 
this paper proposes \textit{PrometheusFree}, an optical neural network framework that is capable of concurrent detection of laser fault injection attacks. 
This paper first presents an illustrative threat of laser fault injection attacks on SPAAs, capable of subjecting the optical neural network to misclassifications. The threat then is addressed in this paper by developing techniques for concurrent detection of the laser fault injection attacks.
Furthermore, this paper introduces a novel application of Wavelength Division Perturbation (WDP) technique where wavelength-dependent Vector Matrix Multiplication (VMM) results are utilized to boost fault attack detection accuracy. Simulation results show that \textit{PrometheusFree} achieves over 96\% attack-caused misprediction recall as the use of the WDP technique squashes the attack success rate by 38.6\% on average. Compared with prior art, \textit{PrometheusFree} limits the average attack success ratio to 0.019, yielding a 95.3\% reduction.
The experimental results confirm the superiority of the concurrent detection and the boost in attack detection abilities imparted by the WDP approaches.
\end{abstract}


\section{Introduction}
Artificial Intelligence (AI)-based applications play an important role
in the rapid development of our super smart society.
The continuous downscaling of CMOS transistors has been exploited heretofore to satisfy the
ever-increasing demands for computing efficiency in our society.
However, improvements in CMOS circuit latency
have saturated
since parasitic delay has become dominant in advanced process technologies \cite{irds2023}.
Optical circuits have emerged as a promising approach for
resolving the latency wall of CMOS circuits.
Traditionally, optical communication technologies have been utilized for 
long-distance efficient communication. Recent advancements in silicon photonics
have enabled ever-shorter optical network-on-chip applications \cite{onoc}.
Recently, optical computing by silicon photonics has been extensively
studied to boost up computational efficiency.
Motivated largely by the needs of the rapid emerging AI applications,
various types of Optical Neural Networks (ONNs)
have been widely studied \cite{Fu2024}. Recently, many start-up companies such as Lightmatter \cite{lightmatter},
Luminous, LightOn, Lightelligence and Q.ANT have released commercial silicon photonics-based AI accelerators (SPAAs),
signaling their imminent widespread adoption.
This development foreshadows optics-based edge computing 
with optical computing devices  employed as midrange edge servers along with mobile devices
to meet the ever-increasing demand for AI performance as pointed out in \cite{kitayama2019}. In the research direction of photonic computing, \cite{edgephotonics} presents a proof-of-concept demonstration claiming DNNs can be executed on photonic devices on both edge devices and servers.

The question of mitigating the vulnerabilities of ONNs to physical security attacks, however, has received attention only recently, putting the viability of the adoption of ONNs as a supporting technology for AI applications into sharp question.
A fundamental challenge that would need to be addressed to solve the issues raised is the exposure to attackers that can access SPAAs for malicious purposes.
For example, Deep Neural Networks (DNNs) are
widely used in mission-critical and security-sensitive systems such as autonomous driving systems.
In these contexts, the frequently presumed physical protection falls short as unprotected edge servers can be easily accessed by an attacker leading to widespread client malfunctions possibly.
The injection of even 
a tiny fault could cause the DNN to misclassify the input data
 \cite{dnn_fi},
leading to serious accidents in  mission-critical systems. 
A further challenge is the sheer size of 
optical circuit elements, typically several orders of magnitude larger than CMOS ones,
motivating attackers to access SPAAs for malicious purposes due to the facile and precise identification and manipulation of the target elements.

This paper promises to unbound the shackles of security vulnerability of ONNs by introducing \textit{PrometheusFree}, an optical neural network framework that can perform concurrent detection of laser fault injection attacks.
Given the urgency and criticality of SPAA attack vulnerability, 
this paper targets the thermal vulnerability of SPAAs, a viable laser fault injection attack.
This paper goes on to introduce a capable defense 
to protect against the attack vulnerability identified. 
The concrete contributions of this paper are summarized as follows:
\begin{enumerate}
\item This paper introduces a threat of laser fault injection attacks on SPAAs.
The measurement result shows
even a single laser beam can inject a $\pi$ phase shift
into an SPAA, which is sufficient to cause ONN misprediction.
\item  
This paper proposes \textit{PrometheusFree}, featuring a concurrent monitoring technique
of laser fault injection attacks, yet incurring no accuracy degradation.
A checksum added in the model layers monitors abnormal phase shifts
caused by laser fault injection attacks in optical circuits.
\item
This paper introduces a concept of Wavelength Division Perturbation (WDP)
 where wavelength-dependent Vector Matrix Multiplication (VMM) results are utilized for detection.
This technique amplifies the effect of abnormal phase shifts in VMMs,
thereby increasing detection accuracy without degrading ONNs' latency, while incurring minimal hardware overhead.
\item A simulation environment of fault injection attacks targeted in this paper is developed.
Simulation results show \textit{PrometheusFree} achieves on average over 96\%
attack-caused misprediction recall.
\end{enumerate}
The rest of this paper is organized as follows.
Section~\ref{sec:motivation} presents motivations and related works. 
In Section~\ref{sec:framework}, \textit{PrometheusFree} is proposed.
Section~\ref{sec:results} presents its evaluation results.
Section~\ref{sec:conclusion} concludes this paper.

\section{Motivational Example and Related Work}
\label{sec:motivation}
\subsection{Preliminaries: Optical Neural Networks}
\begin{figure}[b!]
  \centering
  \includegraphics[scale=1.0]{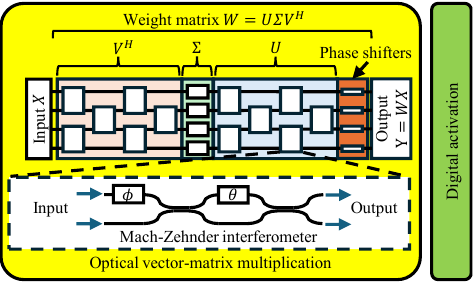}
  \caption{Overview of SPAA architecture using MZI-VMM.}
  \label{fig:mzi}
\end{figure}
Fig.~\ref{fig:mzi} shows an overview of a Mach-Zehnder Interferometer (MZI)-based SPAA architecture \cite{Shen2017}.
It consists of an optical VMM part and a digital activation part
which perform linear transformations and non-linear transformations, respectively.
The digital activation part is processed by a CMOS coprocessor.

For the VMM part,
this paper utilizes an MZI-array Vector-Matrix Multiplier (MZI-VMM) as a representative of SPAAs. 
The weight matrix $W$ is decomposed into unitary matrices $U$ and $V^{H}$, diagonal matrix $\Sigma$, and phase shifts by the Singular Value Decomposition (SVD).
The diagonal matrix $\Sigma$ consists of attenuators or amplifiers.
The Clements mesh~\cite{Clements2016} is employed for constructing $U$ and $V^{H}$. 
The optical circuits corresponding to the Clements mesh
are composed of multiple MZIs.
An MZI consists of two phase shifters and two 3~dB couplers. 
An ideal MZI realizes an arbitrary $2 \times 2$ unitary matrix by continuously tuning $\phi$ and $\theta$.
By tuning $\phi$'s and $\theta$'s of MZIs in an MZI-VMM,
any VMM can be achieved.

\subsection{Motivational Example: Laser Fault Injection Attacks}
Fault Injection Attacks (FIAs) are serious physical attacks for inducing misprediction of DNN-based applications.
DNNs are widely used in mission-critical
and security-sensitive systems such
as autonomous driving systems.
If attackers can induce misprediction
in DNNs in such systems, these attacks
can end up having serious repercussions on the systems.
For traditional CMOS AI accelerators, attackers tamper
with parameters stored in memories
\cite{dnn_fi} or with activation results \cite{dnn_laser_fi}
using FIAs.

\begin{figure}[b!]
  \centering
  \includegraphics[width=8cm]{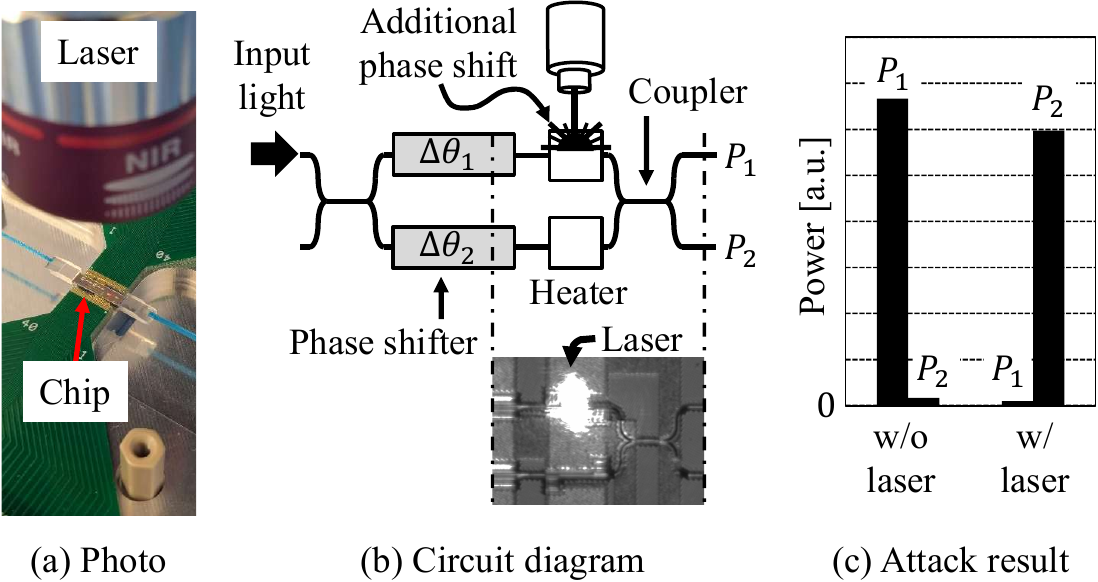}
  \caption{Laser fault injection attack on silicon photonics devices.}
  \label{fig:attack2}
\end{figure}
This paper proposes a novel threat of laser fault injection attacks on optical circuits.
A laser source mounted over a fabricated silicon photonics chip shown in Fig.~\ref{fig:attack2}a 
noninvasively injects an additional phase into an optical circuit.
Fig.~\ref{fig:attack2}b shows a circuit diagram of a
portion of an MZI 
(an MZI corresponding to the phase shifter in Fig.~\ref{fig:mzi} yet without a $\phi$).
Phase shifters based on PIN diodes are used to modulate optical phases (``$\theta$" in Fig.~\ref{fig:mzi}). Generally, optical circuits employ heaters
near couplers to compensate nonideal phase fluctuation introduced by process and environmental variations. When attackers irradiate a laser beam
with a 1425~nm wavelength onto a metal plate of a heater, its temperature locally increases. 
Note that the spot diameter of the laser beam is comparable to the heater size.
An additional phase shift is thus injected by the thermo-optic effect.
This is equivalent to injecting a phase error to a nearby phase shifter.
Fig.~\ref{fig:attack2}c shows a measurement result.
Without a laser beam, the input light goes to the $P_1$ port.
However, with a laser beam, the input light passes to the opposite port ($P_2$).
This is equivalent to injection of a $\pi$ phase shift, the most aggressive phase shift, into a phase shifter.
If the attacker irradiates a laser beam into the heater
on the opposite arm, a negative phase shift can be injected.
This paper numerically confirms in Section~\ref{sec:4} 
that ONN misprediction can be induced by a single laser fault injection only.

Emblematic of the analog optical computation paradigm, ONNs impose special challenges, precluding the use 
of numerous defences that can be employed in 
digital CMOS circuits to detect attacks, 
such as the comprehensive testing of the circuits \cite{fi_testing}, 
the use of fault-tolerant redundant data formats \cite{fi_redundant}
and the utilization of sensors for FIAs \cite{8474958} as countermeasures for the proposed laser fault injection attacks.
The application of such digital methods is highly challenging for analog SPAAs.
The redundant data formats no longer work well since
the proposed laser fault injection attacks directly tamper with phase shifts of the optical devices.
The use of conventional CMOS-based FIA sensor techniques
for non-CMOS optics-based elements poses formidable challenges,
precluding their facile use as a countermeasure as well.
Another physical-level countermeasure would be the use of 
a metal shield over CMOS transistors by utilizing metal interconnections to protect chips from laser beams.
However, the laser beam can be bypassed from the backside
of the CMOS chip~\cite{backside}.
Silicon photonic fabrication technologies are based on CMOS fabrication technologies.
The vulnerability to laser FIAs thus still remains an open question in need of resolution for silicon photonic chips as well.

\subsection{Related Work on Tamper-resilient SPAAs}
\label{sec:related}

While software-based attacks such as adversarial attacks~\cite{adversarial} on SPAAs have been discussed, the concerns of hardware security have also recently received attention. In general, two major units in SPAAs are targeted: CMOS controllers and optical processing units. 
SPAAs typically employ electronic controllers
to modulate optical circuits.
Their modulation configurations are stored in CMOS memories.

For these CMOS controllers, \cite{faoptics} introduces the phase gradient attacks identifying the optimal abnormal phase
in each phase shifter to induce the erroneous inference results.
The memory values can be tampered with by
bit-flip attacks as highlighted in~\cite{unlikely}. 
In addition, \cite{serlos} points out a threat of Hardware Trojans (HTs) which are maliciously inserted to the electronic controllers by untrusted entities.
HTs manipulate the propagation direction of the optical signals on MZIs, resulting in erroneous outputs.
Malicious electro-optic circuits interfere with optical elements and lead to erroneous inference results.
A similar situation is discussed in~\cite{safelight}.
For optical computing units, security vulnerabilities caused by thermal crosstalk are pointed out in \cite{serlos}. 
Malicious optical circuits can increase their own temperature, thus elevating the temperature of the victim optical circuits near them through thermal crosstalk, resulting in erroneous outputs. 

To tackle these vulnerabilities on SPAAs, \cite{serlos,unlikely,safelight} propose countermeasures featuring detection and error-recovery mechanisms. 
However, it is hard to use these countermeasures directly to detect laser fault injection attacks on optical circuits.
Ref.~\cite{serlos} proposes a simple attack detection method
where test signals are fed into the ONNs and the output signals are then compared with the expected outputs. This testing method performed at specific intervals achieves 100\% anomaly detection in the various scenarios, whereas it results in the degradation of the SPAAs' potential performance
since SPAAs end up processing test signals frequently in order to monitor the laser fault injection attacks. Moreover, SPAAs are vulnerable if the attackers can distinguish
the normal inference phase and the testing phase.
The laser fault injection attacks are no longer detectable if attackers can only tamper with SPAAs operating in the normal inference phase.
Ref.~\cite{safelight} introduces methods for improving robustness of inference operations with little performance overhead even if HTs tamper with internal signals. 
While this method can be applied as a countermeasure for laser fault injection attacks shown in Section~\ref{sec:latency},
the method falls short of fully compensating for the accuracy degradation
if attackers inject aggressive errors such as laser fault injection attacks.
Ref.~\cite{unlikely} proposes a recovery method and a robust number representation method against bit-flip attacks on CMOS controllers. This method utilizes a checksum validation method to detect the bit-flip attacks and a number representation with low sensitivity to the attacks. Since the targeted
laser fault injection attacks directly induce analog perturbations
in optical circuits, an alternative approach is required to adopt the method outlined in Ref.~\cite{unlikely} as a countermeasure against the attacks.

\begin{figure}[b!]
    \centering
    \includegraphics[scale=1.0]{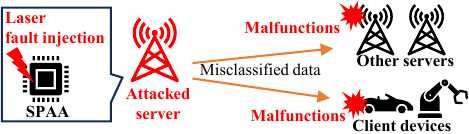}
    \caption{Threat model.}
    \label{fig:scenario}
\end{figure}
\section{PrometheusFree Framework}
\label{sec:framework}
\subsection{Threat Model}
This paper assumes the attackers' motivation is to cause malfunctions in edge photonic computing systems.
Fig.~\ref{fig:scenario} depicts the attack scenario. Considering mission-critical systems as the specific target,
attackers attempt to force DNNs to misclassify
the input data into random wrong categories
thus inducing serious damage to the systems.
Attackers access SPAAs implemented on an edge server and 
locally inject phase shift errors into arbitral
phase shifters in SPAAs.
Specifically, the attacker injects phase errors
into phase shifters in an SPAA ($\phi$'s and $\theta$'s 
in Fig.~\ref{fig:mzi}) that remain operational throughout the entire inference process for
each input data. 
\subsection{PrometheusFree Overview}
This paper outlines \textit{PrometheusFree}, a neural network framework that is capable of
concurrently detecting laser fault injection attacks.
Fig.~\ref{fig:overview} overviews the framework\footnote{We have used colors consistently across Figs.~\ref{fig:overview}, \ref{fig:basearch}, \ref{fig:layerdesign} and \ref{fig:proposed_detail}
 to denote the identical components at various levels of implementation.}.
This framework features the following three parts:
(1) ``Training,'' (2) ``Weight conversion'' and (3) ``Inference.''
(1) ``Training" features a training flow considering the concurrent detection technique
inspired by the balanced output partition technique~\cite{balanced_output}.
This technique is outlined in Section~\ref{sec:PrometheusFree}.
In (2) ``Weight conversion," weight values in ONNs are converted to phase shift values of MZIs ($\phi$'s and $\theta$'s in Fig.~\ref{fig:mzi}).
(3) ``Inference" includes inference frameworks including
the balanced output partition technique and a fault injection simulator.
This paper proposes a concept of \textit{Wavelength Division Perturbation (WDP)}
where abnormal phase shifts are amplified by feeding multiple lights with different wavelengths.
\textit{PrometheusFree} also features a simulation environment presented in Section~\ref{sec:simulation} below.
\begin{figure}[t!]
    \centering
    \includegraphics[scale=1.0]{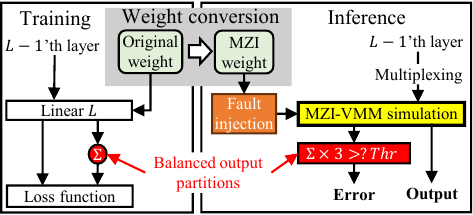}
    \caption{\textit{PrometheusFree} overview.}
    \label{fig:overview}
\end{figure}
\subsection{Simulation Framework}
\label{sec:simulation}
The proposed framework converts the weight values of the target neural network
into phase shift values of MZIs ($\phi$'s and $\theta$'s in Fig.~\ref{fig:mzi}).
The simulator then obtains optical characteristics (amplitudes and phases) of
output lights in each layer considering laser fault injection attacks.
The flow of the simulation framework in Fig.~\ref{fig:overview} is as follows:
(1) Model training without MZI-VMM simulation (numerical optimization of the weight matrix $W$),
(2) Conversion of the weights for the MZI-VMM (decomposition to $\phi$'s and $\theta$'s in Fig.~\ref{fig:mzi}), and 
(3) Inference operation with the MZI-VMM simulator. 
In (2) and (3), this paper targets the optical multiplier architecture illustrated
in Fig.~\ref{fig:basearch}.
This architecture has $N_{\rm VMM}$ MZI-VMMs. Each MZI-VMM is composed of a set of MZI-VMM of matrix size $S_{\rm VMM}$ and homodyne detectors.
It numerically calculates the VMM results $\lceil R/S_{\rm VMM}\rceil \cdot\lceil C/S_{\rm VMM}\rceil$ times, where $R$ and $C$ are the numbers
of rows and columns of the weight matrix $W$, respectively.
The multiplication results are accumulated digitally at the accumulation unit 
to obtain the $R\times C$-size VMM result.
The strength of the input lasers and the output signals
are quantized by the DAC and the ADC, respectively.
Similarly, the signals to set weights are quantized.

The simulation architecture in this paper is inspired by the neural network architecture proposed in \textit{AnalogVNN}~\cite{analogvnn}, a simulation framework
for optical neural networks.
Fig.~\ref{fig:analoglayer} illustrates the \textit{AnalogVNN} framework.
\textit{AnalogVNN} comprises a Digital-Analog Conversion (``DAC'') layer,
a linear operation (``Conv/FC'') layer, and an Analog-Digital Conversion (``ADC'') layer.
The ``DAC'' and the ``ADC'' layers consist of normalization, precision reduction, and noise addition.
The latter two layers quantize the inputs and add Gaussian noise to them, respectively.
In ``Conv/FC," VMM for a target layer ($L$'th layer in this case) is calculated numerically.
Fig.~\ref{fig:modeling} shows the layer design.
We replace the ``Conv/FC'' in Fig.~\ref{fig:analoglayer} with
the in-house MZI-VMM simulator and omit the clamp layers to allow full-range operands.
This simulator ignores manufacturing deviations and executes multiple MZI-VMM simulations in parallel.


\begin{figure}[t!]
  \centering
  \includegraphics[scale=1.0]{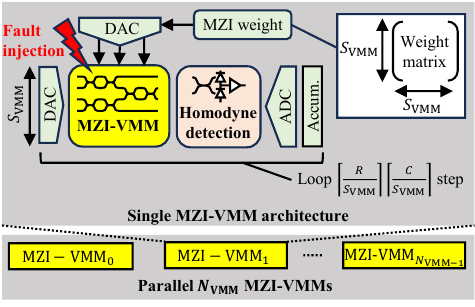}
  \caption{The target SPAA architecture.}
  \label{fig:basearch}
\end{figure}
\begin{figure}[b!]
\begin{minipage}{\linewidth}
    \centering
    \includegraphics[scale=1.0]{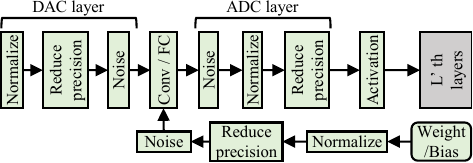}
    \subcaption{An analog layer design.}
    \label{fig:analoglayer}
\end{minipage}\\
\begin{minipage}{\linewidth}
  \centering
  \includegraphics[scale=1.0]{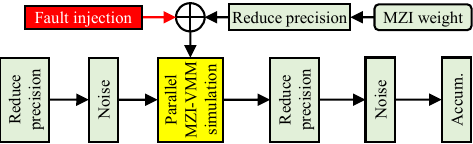}
  \subcaption{The layer design of the SPAA architecture.}
  \label{fig:modeling}
\end{minipage}
\caption{The layer design of \textit{AnalogVNN} and the proposed simulator.}
\label{fig:layerdesign}
\end{figure}

\subsection{ONN for Concurrent Detection of Laser Fault Injection Attacks}
\label{sec:PrometheusFree}
This paper utilizes the balanced output partition technique proposed in \cite{balanced_output} for concurrently detecting laser fault injection attacks.
While originally designed to concurrently detect hardware faults such
as single event upsets in neural networks employed in safety-critical digital systems, 
this paper explores the efficacy of this technique for detecting laser fault injection attacks on analog SPAAs.
Balanced output partitions are designed to embed a balance checksum in deep neural networks.
This technique adds an additional node with fixed weights
in each layer as shown in Fig.~\ref{fig:checksum}.
Half of the edges have a positive unit weight value while
the other half have a negative unit weight value. The weights are trained to minimize its checksum toward zero. 
In the inference phase, 
it is deemed that the ONN is being attacked
if the checksum result exceeds a predetermined threshold.
The lightweight checksum operation for this additional node
is performed by a CMOS coprocessor. No additional hardware is required for optical circuits.
Our thorough evaluation shows
the latency time for the checksum calculation to be considerably smaller
compared to the inference latency time of ONNs under practical device parameters.
\begin{figure}[tb!]
\begin{minipage}{\linewidth}
\centering
  \includegraphics[scale=1.0]{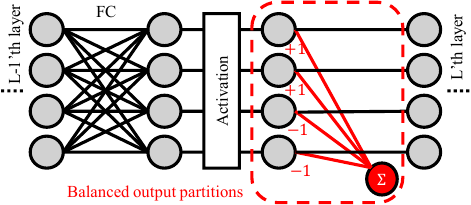}
  \subcaption{Balanced output partitions.}
  \label{fig:checksum}
\end{minipage}\\
\begin{minipage}{\linewidth}
\centering
  \includegraphics[scale=1.0]{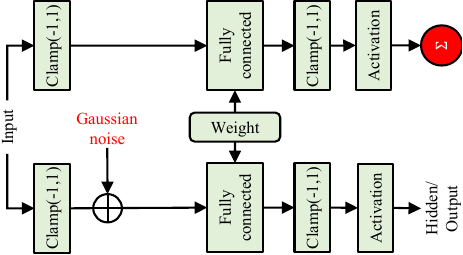}
  \subcaption{Training architecture.}
  \label{fig:training}
\end{minipage}\\
\begin{minipage}{\linewidth}
  \centering
  \includegraphics[scale=1.0]{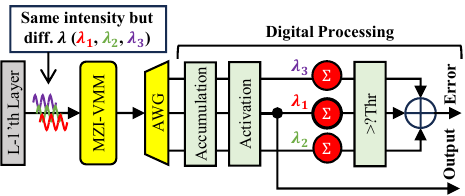}
  \subcaption{Inference and runtime detection architecture ($\lambda$: Wavelength).}
  \label{fig:inference} 
\end{minipage}
\caption{\textit{PrometheusFree} training/inference architecture.}
\label{fig:proposed_detail}
\end{figure}

In this paper, we incorporate this method into an analog layer
as shown in Fig.~\ref{fig:training}, which shows the training architecture. The training architecture has two branches. While the upper branch is used to train the checksums, the lower counterpart focuses on enhancing the inference capability and the robustness of the model. 
We omit the ``Reduce precision" and the ``Noise" layers
in the original ``DAC" and ``ADC" layers in Fig.~\ref{fig:analoglayer} to boost the training speed, except for the noise layer on the input side of the lower branch. The ``Noise'' layer on the lower branch is retained to give the model robustness, as inspired by \textit{Safelight}~\cite{safelight}. 

\subsection{WDP: Wavelength Division Perturbation}
\label{sec:wdm}
The balanced output partitions do not detect a laser fault injection attack
if the attack has no material impact on the checksum results.
For example, if the attack increases the 
sums of the two halves 
of the output values in Fig.~\ref{fig:checksum}  by the same amount,
the checksum result remains constant, which leads to misdetection (false negative).
To increase the detection capability, a concept of \textit{Wavelength Division Perturbation (WDP)} is proposed.
The key point is that the actual modulation amounts of $\phi$ and $\theta$ in Fig.~\ref{fig:mzi} depend on the light wavelength.
We input light signals with identical intensities but different wavelengths to ONNs.
As a result, the weight matrix $W$
is ``perturbed" in a wavelength-dependent manner.
The intensities of the output lights with multiple wavelengths are thus different.
This changes the balance at the checksum node,
thereby amplifying the effect of the abnormal phase shift even if
the attack does not impact the checksum results on the original wavelength.
\textit{WDP} allows \textit{PrometheusFree} not only
to perform inference in a target wavelength but also
to detect attacks in all the wavelengths concurrently, yet has no impact on
the operating speed of the main inference operation.
In order to obtain the checksum results individually, the number of checksum nodes
in the output layer (``$\Sigma$" node in Fig.~\ref{fig:checksum}) is set to
the number of wavelengths. 

Fig.~\ref{fig:inference} shows a case where three wavelengths
($\lambda_1$, $\lambda_2$ and $\lambda_3$) and three checksum nodes (``$\Sigma$"s) are used.
The optical signal with $\lambda_1$ provides not only a VMM result
but also a checksum result. The other two input signals with $\lambda_2$ and $\lambda_3$
have the same intensities as the original $\lambda_1$ signal.
They are concurrently used solely for obtaining the checksum results.
Since $W$ elements differ by wavelengths,
the three checksum results are different, thereby boosting detectability.
If at least one of the checksum results exceeds a predetermined threshold,
it is judged that the ONN has been attacked.
The threshold is set to an appropriate nonzero number to reduce the misdetection (false positive) rate that a too low threshold, such as 0, would otherwise result in.

To incorporate the WDP effect, the proposed simulator assumes that MZIs have the wavelength dependencies of both the directional couplers and the phase shifters.
This simulator incorporates the wavelength dependencies based on the simulation results of silicon photonics devices.

\section{Results and Discussion} \label{sec:4}
\label{sec:results}
\subsection{Setup}
\label{sec:setup}
\subsubsection{Workload}\label{sec:workload}
We apply the proposed method to MLP-Mixer~\cite{10.5555/3540261.3542118} with the German Trafﬁc Sign Recognition Benchmark Dataset (GTSRB)~\cite{STALLKAMP2012323}. 
 Fig.~\ref{fig:mlp-mixer} and Table~\ref{tab:modelsetup} show the model architecture and the parameter definitions, respectively. Each $32\times32$ image with 3 channels is split into 64 patches with $4\times4$ pixels. The linear projection with the output size 63, whose size is used as input and output in the mixer layers, is applied to each patch. The CLS-token which has 63 dimensions is appended as the patch.
The hidden size of token mixing and channel mixing in the mixer layers is 127.
After passing through four mixer layers, the image is classified into 43 labels by an additional fully connected layer inputting the CLS-token.
\begin{figure}[!t]
    \centering
    \includegraphics[scale=1.0]{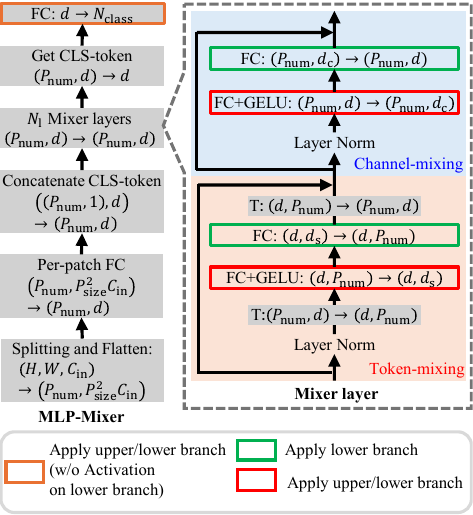}
    \caption{MLP-Mixer architecture.}
    \label{fig:mlp-mixer}
\end{figure}
\begin{table}[!b]
    \centering
    \caption{Notation and parameters.}
    \label{tab:modelsetup}
    \begin{tabular}{c|c|c}
        \hline
         Notation&Definition&Parameter  \\
         \hline
         \hline
         $C_{\rm in}$&The number of channels of input images&3\\
         \hline
		 $P_{\rm num}$&The number of Patches&64\\
		 \hline
		 $P_{\rm size}$&The size of Patch&4\\
		 \hline
		 $N_{\rm l}$&The number of mixer layers&4\\
		 \hline
		 $d$&The input and output size of mixer layer&63\\
		 \hline
		 $d_{\rm c}$&The hidden size in channel--mixing&127\\
		 \hline
		 $d_{\rm s}$&The hidden size in token--mixing&127\\		 
         \hline
		 $N_{\rm class}$&The number of classes in the dataset&43\\
		 \hline
    \end{tabular}
\end{table}
We apply the proposed architecture as shown in Fig.~\ref{fig:mlp-mixer}. The upper and the lower branches in Fig.~\ref{fig:training} are applied to the layers enclosed by red solid line.
We also apply the upper and the lower branches of the training architecture to the layer framed in orange, where the lower branch does not include an activation function. On the other hand, only the lower branch is applied to the layers framed in green.
The Leaky ReLU and the GELU functions are adopted as the activation functions of the upper part and the lower part of Fig.~\ref{fig:training} (``Activation"s near ``$\Sigma$" and ``Hidden/Output"), respectively. 
Gaussian noise 
is added to input signals of the lower part of Fig. \ref{fig:training} in order to incorporate noise robustness. 
For maximizing inference accuracy, the standard deviation of the Gaussian noise is swept and set to 0.1 as the best parameter.
On the device setup shown in Sect.~\ref{sec:devicesetup}, the model achieves 97.34\% recognition accuracy.

\subsubsection{Device}
\label{sec:devicesetup}
The device setup is shown in Table~\ref{tab:devicesetup}. 
We add Gaussian noise with a 0.02 standard deviation to the input/output signals. 
A 1556~nm wavelength is used
as the default wavelength for inference ($\lambda_{1}$ in Fig.~\ref{fig:inference}).
$N_{\rm VMM}$ MZI-VMMs are used so that a $256\times256$ sized matrix multiplication
can be performed by the $N_{\rm VMM}$ MZI-VMMs in parallel.
This paper explores the optimal combination of $N_{\rm VMM}$ and $S_{\rm VMM}$
in terms of area and latency. The area is estimated with the hardware area estimator in \cite{10476418}.
The latency is approximated by the number of times MZI-VMMs are used.
The area evaluation shows that the latency is essentially constant up to $S_{\rm VMM}=64$ while increasing sharply when $S_{\rm VMM}$ is 128.
On the other hand, the area decreases as $S_{\rm VMM}$ increases.
As a tradeoff point, this paper utilizes the
configuration where the 16 MZI-VMMs with a $64\times64$ size are used in parallel to perform $256\times256$ VMM operations (i.e., $S_{\rm VMM}=64$ and $N_{\rm VMM}=16$).
\begin{table}[!t]
    \centering
    \caption{MZI-VMM parameters.}
    \label{tab:devicesetup}
    \begin{tabular}{c|c|c}
        \hline
         \multicolumn{2}{c|}{Settings}&Parameters  \\
         \hline
         \hline
         Laser Sources /&Reduce precision & 16 bit \\
         Homodyne Detectors&Noise deviation&0.02\\
         \hline
         \multirow{3}{*}{MZI-VMM Simulator}
         &Reduce precision&16 bit\\
         &Default wavelength&1556 nm\\
        &Overall matrix size & $256\times 256$ \\
         \hline
    \end{tabular}
\end{table}
\subsubsection{Detection and Attack Situation Setup}
The wavelengths for WDP ($\lambda_{2}$ and $\lambda_{3}$) are set to $1550~{\rm nm}-\Delta\lambda$ and $1560~{\rm nm}+\Delta\lambda$, respectively.
We evaluate the detection accuracy by
sweeping $\Delta\lambda$ from 0~nm to 40~nm. 
The thresholds in \textit{PrometheusFree} are determined so that the misdetection rate hovers around 10\% without attacks. 
In an attack situation, it is assumed that 1, 3, 5 and 7 phase shifters are attacked in an SPAA throughout the entire inference operation of every sample
by the laser fault injection attacks. Phase shifts of $\pi/2$, $-\pi/2$ and $\pi$ radian are injected.
The attack points are selected randomly with 1000 attack trials so that an MZI-VMM processing the final layer is subjected to at least one phase shift attack.

\subsection{Detection Result}
\label{sec:detection_result}

This paper evaluates the attack-caused misprediction recall as the detection accuracy and attack success ratio.
A recall metric is defined as $Recall=TP/\left(TP+FN\right)$,
where $TP$ and $FN$ denote the numbers of true positive and false negative samples, respectively. The attack success ratio is the ratio of the number of successful attacks. A successful attack is an attack in which ONNs have been forced to misclassify the input images into random wrong categories by laser fault injection attacks yet 
without being detected by the proposed concurrent detection techniques. The attack success ratio, i.e. the number of the undetected attack-caused mispredictions, 
is formulated as $FN$.

Fig.~\ref{fig:wavelengthdiff} and \ref{fig:fasuccess_wdp} show the simulation results for the misprediction recall and the attack success ratio.
For each ``The number of attack points'' result, the attack success ratio is normalized by the $FN$ value when the WDP technique is not used (hereinafter referred to as ``single wavelength mode"). 
 The configuration of $\Delta\lambda=10~{\rm nm}$ ($\lambda_{2}=1550~\rm nm - \Delta\lambda=1540~{\rm nm}$ and $\lambda_{3}=1560~\rm nm + \Delta\lambda=1570~{\rm nm}$)
 labeled as ``w/ WDP'' in Fig.~\ref{fig:simresults}, which performs the best detectability, showcases the  \textit{PrometheusFree} results.
The single wavelength mode labeled as ``w/o WDP'' in Fig.~\ref{fig:wavelengthdiff} achieves over 94\% average recall. By introducing the WDP, the detection performance improves and achieves over 96\% average recall in each situation. For the attack success ratio, the WDP technique achieves, on average, a 38.6\% reduction as shown in Fig.~\ref{fig:fasuccess_wdp}.
\begin{figure}[tb!]
\begin{minipage}{\linewidth}
\centering
    \includegraphics[scale=1.0]{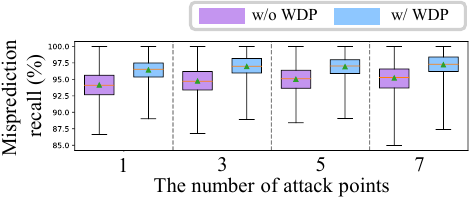}
    \subcaption{A box plot of the attack detection accuracy.}
    \label{fig:wavelengthdiff}
\end{minipage}\\
\begin{minipage}{\linewidth}
\centering
  \includegraphics[scale=1.0]{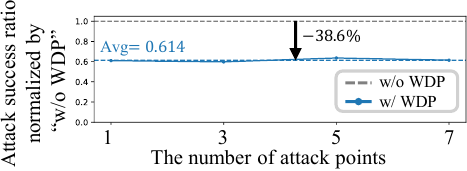}
  \subcaption{Attack success ratio normalized by \textit{PrometheusFree} without WDP.}
  \label{fig:fasuccess_wdp}
\end{minipage}
\caption{Simulation results without WDP and with WDP.}
\label{fig:simresults}
\end{figure}

\subsection{Comparison with Prior Art}\label{sec:latency}
We compare \textit{PrometheusFree} with the baseline, where tamper-resilient techniques are not applied, and the \textit{Safelight}~\cite{safelight} approach.
As mentioned in Section~\ref{sec:related}, \textit{SerIOS}~\cite{serlos}, a testing-based method, is a strong countermeasure to various attack scenarios. Nonetheless, it is difficult to adopt the method since the method leads to throughput degradation of SPAAs. Moreover, the \textit{SerIOS} approach is still vulnerable if attackers can distinguish between test phases and inference phases.
It is also hard to apply \textit{UnlikelyHero}~\cite{unlikely} employing a checksum validation approach to detect attacks on electrical controllers and a number representation with low sensitivity to the attacks. While \textit{UnlikelyHero} is effective against attacks on CMOS memories, it does not consider the scenario
where optical devices are directly attacked.
\textit{Safelight}, which utilizes the Noise-Aware Training (NAT) approach, is an applicable approach as a countermeasure to mitigate laser fault injection attacks. NAT is a training method injecting noise to neural network architectures in a training phase.
In \textit{Safelight}, the accuracy degradation is caused by HTs in electrical controllers for ONNs.
 The effects of their malicious activities are considered as noise.
By injecting noise in the training phase, \textit{Safelight} realizes tamper-resilient ONNs.
This strategy is effective against the laser fault injection attacks as well. 
This paper injects Gaussian noise into the input, the output, and all hidden layers in the model. The standard deviation of the noise is varied
from 0.1 to 0.4. The most robust result of \textit{Safelight} is compared
with the proposed method (\textit{PrometheusFree}).

For a fair comparison with the prior art, we extend the definition of the attack success ratio.
The attack success ratio for the proposed methods is formulated as $FN$ as described in Section~\ref{sec:detection_result}. The ratios for ``Baseline"
and ``Safelight" are the number of successful attacks in which ONNs are
forced to misclassify the input images into random wrong
categories by the laser fault injection attacks.
Fig.~\ref{fig:comparison} shows the comparison results. The attack success ratio is normalized by the values of ``Baseline.'' 
The \textit{Safelight} approach reduces the attack success ratio by  60.2\% on average. In contrast, the proposed method with WDP squashes the average attack successes by over 98\%. Compared with \textit{Safelight} results, \textit{PrometheusFree} ``w/o WDP'' and ``w/ WDP'' achieves 92.3\% and 95.3\% reduction on average, respectively. 
\begin{figure}[tb!]
    \centering
    \includegraphics[scale=1.0]{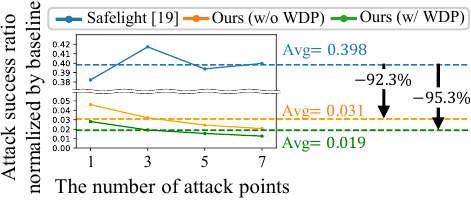}
    \caption{Comparison with prior art by attack success ratio.}
    \label{fig:comparison}
\end{figure}

\section{Conclusion}
\label{sec:conclusion}
This paper proposes \textit{PrometheusFree}, an optical neural network framework
enabling not only inference operations but also concurrent detection of laser fault injection attacks on optical circuits. 
\textit{PrometheusFree} features the concurrent detection
of the laser fault injection attacks as well as providing a simulation environment capable of 
considering the laser fault injection attacks.
Furthermore, \textit{PrometheusFree} employs WDP which fully exploits the nature of light and boosts the detection rate without degrading the operating speed.
The evaluation results show that 
the proposed method achieves over 96\% 
attack-caused average misprediction recall and squashes the attack success ratio by over 98\% compared with the baseline.


\section*{Acknowledgment}
This work was supported in part by JSPS KAKENHI (Grant Numbers 23H04804 and 22H05192), JST FOREST (Grant Number JPMJFR232Q), and JST ASPIRE (Grant Number JPMJAP2429). This work was partly achieved through the use of SQUID at D3 Center, The University of Osaka.
This work was also supported through the activities of VDEC, d.lab, The University of Tokyo, in collaboration with Cadence Design Systems.
The authors thank Prof. Michihiro Shintani of Kyoto Institute of Technology and Prof. Masanori Hashimoto of Kyoto University for valuable advices on this research.

\bibliographystyle{IEEEtran}
\bibliography{ref}

@misc{irds2023,
  title        = {{International roadmap of devices and systems (IRDS). More Moore. 2023.}},
  note         = {{Available online: \url{https://irds.ieee.org/editions/2023/20-roadmap-2023-edition/130-irds%E2%84%A2-2023-more-moore} (Accessed on Sep. 18, 2024.)}}
}

@article{
edgephotonics,
author = {Alexander Sludds  and Saumil Bandyopadhyay  and Zaijun Chen  and Zhizhen Zhong  and Jared Cochrane  and Liane Bernstein  and Darius Bunandar  and P. Ben Dixon  and Scott A. Hamilton  and Matthew Streshinsky  and Ari Novack  and Tom Baehr-Jones  and Michael Hochberg  and Manya Ghobadi  and Ryan Hamerly  and Dirk Englund },
title = {{Delocalized photonic deep learning on the internet’s edge}},
journal = {Science},
volume = {378},
number = {6617},
pages = {270-276},
year = {2022},
doi = {10.1126/science.abq8271},
eprint = {https://www.science.org/doi/pdf/10.1126/science.abq8271}}

@ARTICLE{8474958,
  author={Matsuda, Kohei and Fujii, Tatsuya and Shoji, Natsu and Sugawara, Takeshi and Sakiyama, Kazuo and Hayashi, Yu-Ichi and Nagata, Makoto and Miura, Noriyuki},
  journal={IEEE Journal of Solid-State Circuits}, 
  title={{A 286 F$^2$/Cell distributed bulk-current sensor and secure flush code eraser against laser fault injection attack on cryptographic processor}}, 
  year={2018},
  volume={53},
  number={11},
  pages={3174-3182},
  doi={10.1109/JSSC.2018.2869142}}

@article{onoc,
author = {Wu, Xiaowen and Xu, Jiang and Ye, Yaoyao and Wang, Zhehui and Nikdast, Mahdi and Wang, Xuan},
title = {{SUOR: Sectioned undirectional optical ring for chip multiprocessor}},
year = {2014},
issue_date = {May 2014},
publisher = {Association for Computing Machinery},
address = {New York, NY, USA},
volume = {10},
number = {4},
issn = {1550-4832},
doi = {10.1145/2600072},
journal = {J. Emerg. Technol. Comput. Syst.},
month = {jun},
articleno = {29},
numpages = {25}
}

@Article{Shen2017,
author={Shen, Yichen
and Harris, Nicholas C.
and Skirlo, Scott
and Prabhu, Mihika
and Baehr-Jones, Tom
and Hochberg, Michael
and Sun, Xin
and Zhao, Shijie
and Larochelle, Hugo
and Englund, Dirk
and Solja{\v{c}}i{\'{c}}, Marin},
title={{Deep learning with coherent nanophotonic circuits}},
journal={Nature Photonics},
year={2017},
month={Jul},
day={01},
volume={11},
number={7},
pages={441-446},
doi={10.1038/nphoton.2017.93}
}

@INPROCEEDINGS{lightmatter,
  author={Ramey, Carl},
  booktitle={2020 IEEE Hot Chips 32 Symposium (HCS)}, 
  title={{Silicon photonics for artificial intelligence acceleration : HotChips 32}}, 
  year={2020},
  volume={},
  number={},
  pages={1-26},
  doi={10.1109/HCS49909.2020.9220525}
}

@article{kitayama2019,
    author = {Kitayama, Kenichi and Notomi, Masaya and Naruse, Makoto and Inoue, Koji and Kawakami, Satoshi and Uchida, Atsushi},
    title = "{{Novel frontier of photonics for data processing--photonic accelerator}}",
    journal = {APL Photonics},
    volume = {4},
    number = {9},
    pages = {090901},
    year = {2019},
    month = {09},
    issn = {2378-0967},
    doi = {10.1063/1.5108912}
}

@INPROCEEDINGS{dnn_fi,
  author={Liu, Yannan and Wei, Lingxiao and Luo, Bo and Xu, Qiang},
  booktitle={2017 IEEE/ACM International Conference on Computer-Aided Design (ICCAD)}, 
  title={{Fault injection attack on deep neural network}}, 
  year={2017},
  volume={},
  number={},
  pages={131-138},
  doi={10.1109/ICCAD.2017.8203770}}

@inproceedings{dnn_laser_fi,
author = {Breier, Jakub and Hou, Xiaolu and Jap, Dirmanto and Ma, Lei and Bhasin, Shivam and Liu, Yang},
title = {{Practical fault attack on deep neural networks}},
year = {2018},
isbn = {9781450356930},
doi = {10.1145/3243734.3278519},
booktitle = {Proceedings of the 2018 ACM SIGSAC Conference on Computer and Communications Security (CCS)},
pages = {2204–2206},
numpages = {3},
location = {Toronto, Canada}
}

@ARTICLE{fi_testing,
  author={Su, Fei and Liu, Chunsheng and Stratigopoulos, Haralampos-G.},
  journal={IEEE Design \& Test}, 
  title={{Testability and dependability of AI hardware: Survey, trends, challenges, and perspectives}}, 
  year={2023},
  volume={40},
  number={2},
  pages={8-58},
  doi={10.1109/MDAT.2023.3241116}}

@ARTICLE{fi_redundant,
  author={Liu, Liang and Guo, Yanan and Cheng, Yueqiang and Zhang, Youtao and Yang, Jun},
  journal={IEEE Transactions on Computers}, 
  title={{Generating robust DNN with resistance to bit-flip based adversarial weight attack}}, 
  year={2023},
  volume={72},
  number={2},
  pages={401-413},
  doi={10.1109/TC.2022.3211411}}

@ARTICLE{adversarial,
  author={Jiao, Shuming and Song, Ziwei and Xiang, Shuiying},
  journal={IEEE Journal of Selected Topics in Quantum Electronics}, 
  title={{Adversarial attacks on an optical neural network}}, 
  year={2023},
  volume={29},
  number={2: Optical Computing},
  pages={1-6},
  doi={10.1109/JSTQE.2022.3207056}}

@article{Clements2016,
   author = {William R. Clements and Peter C. Humphreys and Benjamin J. Metcalf and W. Steven Kolthammer and Ian A. Walsmley},
   doi = {10.1364/optica.3.001460},
   issn = {23342536},
   issue = {12},
   journal = {Optica},
   month = {12},
   pages = {1460},
   publisher = {The Optical Society},
   title = {{Optimal design for universal multiport interferometers}},
   volume = {3},
   year = {2016},
}

@INPROCEEDINGS{balanced_output,
  author={Ozen, Elbruz and Orailoglu, Alex},
  booktitle={25th Asia and South Pacific Design Automation Conference (ASP-DAC)}, 
  title={{Concurrent monitoring of operational health in neural networks through balanced output partitions}}, 
  year={2020},
  volume={},
  number={},
  pages={169-174},
  doi={10.1109/ASP-DAC47756.2020.9045662}}

@article{analogvnn,
   author = {Vivswan Shah and Nathan Youngblood},
   doi = {10.1063/5.0134156},
   issue = {2},
   journal = {APL Machine Learning},
   month = {6},
   publisher = {AIP Publishing},
   title = {{AnalogVNN: A fully modular framework for modeling and optimizing photonic neural networks}},
   volume = {1},
   year = {2023},
}

@article{STALLKAMP2012323,
title = {{Man vs. computer: Benchmarking machine learning algorithms for traffic sign recognition}},
journal = {Neural Networks},
volume = {32},
pages = {323-332},
year = {2012},
note = {Selected Papers from IJCNN 2011},
issn = {0893-6080},
doi = {https://doi.org/10.1016/j.neunet.2012.02.016},
author = {J. Stallkamp and M. Schlipsing and J. Salmen and C. Igel}
}

@inproceedings{10.5555/3540261.3542118,
author = {Tolstikhin, Ilya and Houlsby, Neil and Kolesnikov, Alexander and Beyer, Lucas and Zhai, Xiaohua and Unterthiner, Thomas and Yung, Jessica and Steiner, Andreas and Keysers, Daniel and Uszkoreit, Jakob and Lucic, Mario and Dosovitskiy, Alexey},
title = {{MLP}-mixer: an all-{MLP} architecture for vision},
year = {2021},
isbn = {9781713845393},
booktitle = {Proceedings of the 35th International Conference on Neural Information Processing Systems (NIPS)},
articleno = {1857},
numpages = {12}
}

@INPROCEEDINGS{10476418,
  author={Zhu, Hanqing and Gu, Jiaqi and Wang, Hanrui and Jiang, Zixuan and Zhang, Zhekai and Tang, Rongxing and Feng, Chenghao and Han, Song and Chen, Ray T. and Pan, David Z.},
  booktitle={2024 IEEE International Symposium on High-Performance Computer Architecture (HPCA)}, 
  title={{Lightening-transformer: A dynamically-operated optically-interconnected photonic transformer accelerator}}, 
  year={2024},
  volume={},
  number={},
  pages={686-703},
  doi={10.1109/HPCA57654.2024.00059}}

@INPROCEEDINGS{safelight,
  author={Afifi, Salma and Thakkar, Ishan and Pasricha, Sudeep},
  booktitle={Design, Automation \& Test in Europe Conference (DATE)}, 
  title={{SafeLight: Enhancing security in optical convolutional neural network accelerators}}, 
  year={2025},
  volume={},
  number={},
  pages={1-7},
  doi={10.23919/DATE64628.2025.10993109}
}

@ARTICLE{faoptics,
  author={Su, Ye and Jiang, Xiao and Xu, Fang and Ye, Yichen and Chen, Zhuang and Lu, Simi and Liu, Weichen and Xie, Yiyuan},
  journal={IEEE Journal of Selected Topics in Quantum Electronics}, 
  title={{A formal scheme of fault injection on coherent integrated photonic neural networks}}, 
  year={2025},
  volume={31},
  number={3: AI/ML Integrated Opto-electronics},
  pages={1-11},
  doi={10.1109/JSTQE.2024.3493857}}

@inproceedings{serlos,
author = {G. Magalhaes, Felipe and Nikdast, Mahdi and Nicolescu, Gabriela},
title = {{SerIOS: Enhancing hardware security in integrated optoelectronic systems}},
year = {2024},
isbn = {9798400704109},
doi = {10.1145/3625223.3649275},
booktitle = {Proceedings of the 34th International Workshop on Rapid System Prototyping (RSP)},
articleno = {04},
numpages = {7},
location = {Hamburg, Germany}
}

@inproceedings{unlikely,
author = {Lu, Haotian and Yin, Ziang and Bhoumik, Partho and Banerjee, Sanmitra and Chakrabarty, Krishnendu and Gu, Jiaqi},
title = {{The unlikely hero: Nonidealities in analog photonic neural networks as built-in adversarial defenders}},
year = {2025},
isbn = {9798400706356},
doi = {10.1145/3658617.3697771},
booktitle = {Proceedings of the 30th Asia and South Pacific Design Automation Conference (ASP-DAC)},
pages = {1295–1301},
numpages = {7},
location = {Tokyo, Japan}
}

@Article{Fu2024,
author={Fu, Tingzhao
and Zhang, Jianfa
and Sun, Run
and Huang, Yuyao
and Xu, Wei
and Yang, Sigang
and Zhu, Zhihong
and Chen, Hongwei},
title={Optical neural networks: progress and challenges},
journal={Light: Science {\&} Applications},
year={2024},
month={Sep},
day={20},
volume={13},
number={1},
pages={263},
issn={2047-7538},
doi={10.1038/s41377-024-01590-3}
}

@INPROCEEDINGS{backside,
  author={van Woudenberg, Jasper G.J. and Witteman, Marc F. and Menarini, Federico},
  booktitle={2011 Workshop on Fault Diagnosis and Tolerance in Cryptography}, 
  title={{Practical optical fault injection on secure microcontrollers}}, 
  year={2011},
  volume={},
  number={},
  pages={91-99},
  doi={10.1109/FDTC.2011.12}}

\end{document}